# A new method for measuring the neutron lifetime using an *in situ* neutron detector

C. L. Morris[1], E. R. Adamek[2], L. J. Broussard[3], N. B. Callahan[2], S. M. Clayton[1], C. Cude-Woods[4], S. A. Currie[1], X. Ding[6], W. Fox[2], K. P. Hickerson[5], A. T. Holley[7], A. Komives[8], C.-Y. Liu[2], M. Makela[1], R. W. Pattie Jr.[1], J. Ramsey[1], D. J. Salvat[9], A. Saunders[1], S. J. Seestrom[1], E. I. Sharapov[10], S. K. Sjue[1], Z. Tang[1], J. Vanderwerp[2], B. Vogelaar[6], P. L. Walstrom[1], Z. Wang[1], Wanchun Wei[1], J. W. Wexler[1], T. L. Womack[1], A. R. Young[4], and B. A. Zeck[3]

[1]*Los Alamos National Laboratory, Los Alamos, New Mexico 87545, USA*

[2]*Department of Physics, Indiana University, Bloomington, Indiana 47408, USA*

[3]*Oak Ridge National Laboratory, Oak Ridge, TN 37831*

[4]*Triangle Universities Nuclear Laboratory and North Carolina State University, Raleigh, North Carolina 27695, USA*

[5]*California Institute of Technology, Pasadena, California 91125*

[6]*Department of Physics, Virginia Polytechnic Institute and State University, Blacksburg, Virginia 24061, USA*

[7] *Department of Physics,* Tennessee Tech University, Cookeville, TN 38505

[8] *Department of Physics, DePauw University, Greencastle IN 46135-0037*

[9] *Department of Physics, University of Washington, Seattle, WA 98195-1560*

[10]*Joint Institute for Nuclear Research, Dubna, Moscow region, Russia, 141980*

**Abstract:** The neutron lifetime is important in understanding the production of light nuclei in the first minutes after the big bang and it provides basic information on the charged weak current of the standard model of particle physics. Two different methods have been used to measure the neutron lifetime: disappearance measurements using bottled ultracold neutrons and decay rate measurements using neutron beams. The best measurements using these two techniques give results that differ by nearly 4 standard deviations. In this paper we describe a new method for measuring surviving neutrons in neutron lifetime measurements using bottled ultracold neutrons that provides better characterization of systematic uncertainties and enables higher precision than previous measurement techniques. We present results obtained using our method.

## Introduction

Two different techniques have been used to measure the neutron lifetime: by measuring the decay rate in a cold neutron beam using a Penning trap to capture and count resultant protons and by measuring the survival of neutrons after storage using trapped ultracold neutrons (UCN).[1] The most precise measurements from a material bottle (878.5±0.8[2]) and cold beam measurement (887.7±2.2[3]) disagree by 3.9 standard deviations. The probability of both measurements being consistent with the neutron lifetime is about $1\times10^{-4}$.

Because the neutron lifetime controls weak reaction rates for n↔p at freeze out in the early universe and therefore directly affects the $^4$He abundance, the uncertainty in big bang nucleosynthesis (BBN) predictions of the $^4$He abundance is dominated by the uncertainty in the neutron lifetime. [4] Resolving the discrepancy between beam and bottle lifetime results and improving the precision to the sub-one second level is key to improving BBN predictions of primordial elemental abundances. Comparison of the predicted abundances with astrophysical measurements provides additional tests of SM physics.

The unitarity of the Cabbibo-Kobayashi-Maskawa (CKM) matrix provides a test of the standard model sensitive to a host of new physics beyond the standard model.[5] The best test comes from the first row of the CKM matrix because of precise measurements of $V_{ud}$ resulting from an analysis of super allowed nuclear beta decays that dominate the unitarity sum and the uncertainty.[6] Measurements at the level of a few $10^{-4}$ of the neutron lifetime, $\tau_n$, and about $10^{-3}$ in the neutron β asymmetry[7-9], A, can provide a determination of $V_{ud}$ free of the nuclear structure corrections that limit the precision that can be obtained from super-allowed beta decay.[10]

UCN experiments have traditionally used material bottles for neutron storage. In these experiments neutrons are loaded into a bottle and the remaining neutrons are unloaded and counted after a variable storage time. The spectral dependence of neutron up-scatter and absorption leads to spectral softening of the neutron spectrum as a function of storage time. Consequently, the detection efficiency and unloading time become storage-time dependent. Uncertainty in the systematic corrections associated with spectral and phase space evolution form an important contribution to the total lifetime uncertainty in many of the bottle measurements.[2,11-13] Serebrov et al.[2] were able to reduce these effects by using a larger trap to reduce the wall collision rate and lower surface temperature to reduce the loss per wall collision, and have published the smallest uncertainty for the neutron lifetime to date. Still, the largest corrections to the measured lifetime in previous experiments were due to loss on material surfaces. In these experiments this correction was controlled by changing the surface to volume ratio and extrapolating the loss rate to zero, an extrapolation of >5 s for the Serebrov et al. experiment[2] and larger for previous experiments[11,12].

Ezhov *et al*.[14-16] have demonstrated UCN storage in a 20-pole axisymmetric magnetic bottle made of permanent magnets and have reported a preliminary lifetime, $\tau_n$=878.3±1.9 s, in agreement with the Serebrov measurement. The current experiment (UCNτ) aims to reduce systematic uncertainties encountered in these experiments by storing the neutrons in an asymmetric magneto-gravitational trap[17,18] that eliminates wall losses, limits the population of long-lived quasi-bound UCN, and detects the neutrons *in situ* at the end of the storage time. In this paper we describe the *in-situ* detector and demonstrate that shorter counting times can be achieved with this method when compared to previous bottle measurements (viz. the time it takes to "empty" the trap). Further, we investigate the presence of long lived phase space evolution in our trap, a potentially important limit to the precision of 1 s in bottle lifetime measurements.

A cut away view of the trap is shown in Figure 1 and a schematic layout of the beam line is shown in Figure 2. The detector is shown in its lowered position. A storage measurement cycle consists of loading UCN through a removable section at the bottom of the trap (trap door shown in its lowered position in Figure 1), cleaning neutrons with the cleaner lowered to a height of 40 cm above the bottom of the trap, closing the trap door to store neutrons, raising the cleaner and storing neutrons for a variable holding time, and finally lowering the detector (dagger) to count neutrons. The cleaner is a horizontal surface of neutron absorbing material with a small negative potential (In this case $^{10}$B on a ZnS substrate-the same material as the dagger). Neutrons with enough energy to reach the cleaner are expected to eventually cross the cleaner surface and be absorbed.

UCN are provided by the Los Alamos UCN source[19] at the Los Alamos Neutron Science Center (LANSCE) . This is a spallation-driven solid-deuterium UCN source. The 800 MeV proton beam which is used to produce neutrons was only on for the loading period. This results in a low background environment for UCN counting.

A lifetime measurement consists of a sequence of measurements using a short holding time (e.g. 10 s) and a long holding time (e.g. 1410 s), from which the normalized number of UCN is obtained. Typically approximately 15,000 cleaned neutrons are detected in the trap at the short holding time. The statistical precision obtained in the lifetime from a single run pair (~1 h) is about 12 s.

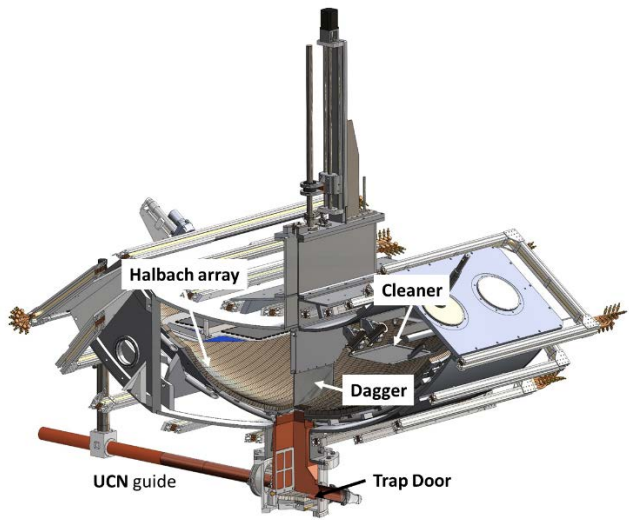

Figure 1) Cross sectional view showing the detector, the actuator and the UCN trap

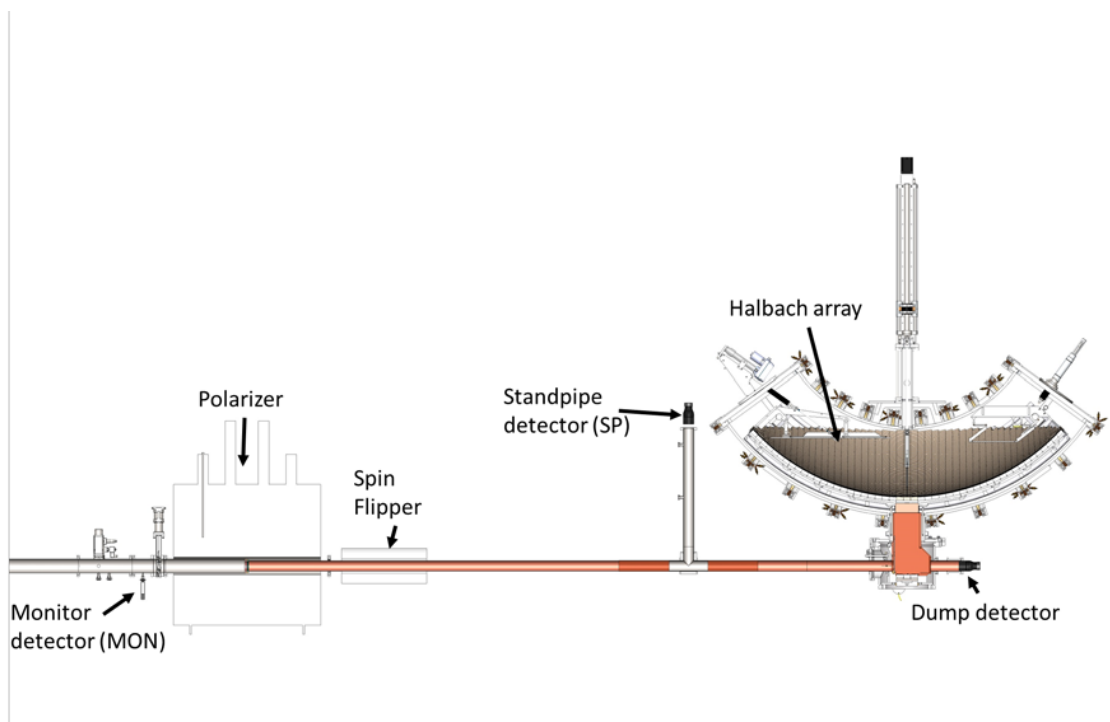

Figure 2) Schematic layout of the UCN beam line showing the monitor detector locations relative to the trap.

## The Detector

UCN were detected using commercial ZnS(Ag) screens [20] coated with 20±5 nm of boron enriched to 95% $^{10}$B that was applied by vacuum evaporation. The maximum energy of neutrons stored in the trap was 40 neV set by the vertical position of the cleaner. The UCN properties of the exposed materials of the detector are listed in Table 1. The Fermi potential for neutrons is given by:

$$V_F = \sum_i \frac{2\pi \hbar^2}{m} N_i a_i, \qquad 1)$$

where $m$ is the neutron mass, $a_i$ is the neutron coherent scattering length, and $N_i$ is the material number density for the $i^{th}$ constituent. The lifetime for neutron absorption in the material is:

$$\tau_A = \frac{1}{\sum_i N_i \sigma_{Ai} v}, \qquad 2)$$

where $\sigma_{Ai}$ are the neutron absorption cross sections and $v$ is the UCN velocity. The cross sections are proportional to the inverse of the neutron velocity ($\sigma_{Ai} \propto 1/v$), and therefore the lifetime (and hence the detection probability) $\tau_A$ is independent of $v$.

As shown in Table 1, all of the materials used in the detector assembly other than $^{10}B$ and acrylic have positive $V_F$ larger than the trap potential, so the UCN are expected to reflect from these materials with an absorption coefficient expected to be in the range of several times $10^{-4}$/reflection, typical of most materials. Further, this material is above the volume of the trap during storage of neutrons. The manufacture of the screen ensures little acrylic is exposed. Scanning electron microscope images the ZnS surface are shown in Ref [21]

The reflection coefficient from the imaginary potential of the $^{10}B$ can be significant[21] by requiring multiple bounces for detection and lengthening the collection time. The effect of the surface roughness of the screen, which may reduce the reflection, has not been quantified. Absorption on the other materials is negligible, even if several bounces are required for a UCN to be absorbed.

Table 1) UCN properties of the detector materials. $V_{Fermi}$ is the surface potential and t is the adsorption time for UCN in the material.

| Material | $V_{Fermi}$ (nV) | Absorption time (ns) |
|---|---|---|
| $^{10}B$ | -3.7 | 8.4 |
| Al | 54.7 | 3.3×10⁵ |
| ZnS | 75.7 | 1.1×10⁵ |
| Acrylic | 27.6 | 2.4×10⁵ |
| Polyimide | 91.2 | 2.8×10⁵ |

Neutrons are captured by the $^{10}B+n \rightarrow \alpha + ^7Li(0\ MeV), ^7Li(0.48\ MeV)$ reaction with its large positive Q-value of 2.79 MeV and 2.31 Mev for the ground state and first excited state of $^7Li$ respectively. The back-to-back correlation of the energetic charged particles ensures at least one will stop in the ZnS(Ag) screen, producing scintillation light. This light is read out using an array of Kuraray Y-10 wave length shifting fibers, WLSF, glued into an ultra-violet transmitting acrylic plate.   The fibers were glued into a set of 1 mm wide, 1 mm deep, 2 mm spaced grooves machined into a 3mm thick plate that was backed by another 3 mm thick plate without grooves.  Alternate fibers were directed into one of two photomultiplier tubes. Photographs of the detector (dagger) are shown in Figure 3.

Some of the light produced in the ZnS(Ag) enters the WLSF, is shifted from blue to green and captured and is transmitted to the phototubes by the fibers. By comparing the light output of the ZnS(Ag) measured with a phototube from a bare screen illuminated with a $^{148}$Gd(3.27 MeV) $\alpha$-particle source with the light output measured for UCN absorption events in the dagger, we estimate the total photon detection efficiency to be 0.9±0.3%.

The dagger was mounted on a linear vacuum feed through that allowed it to be raised and lowered within the UCN trap. In this way UCN could be counted at various heights in the trap. The height resolution is limited by the contour on the bottom of the detector, which was designed to allow the detector to conform to the curved bottom of the trap.

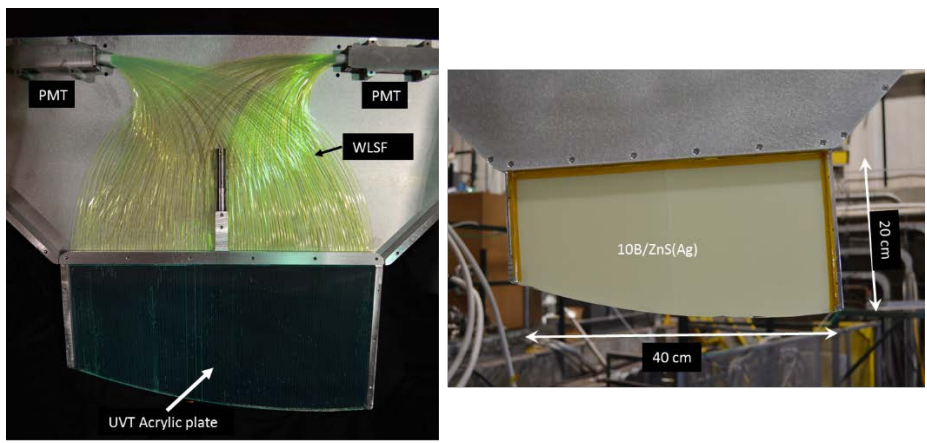

Figure 3 Photographs of the detector. Left) during assembly, showing the phototube housings (PMT), the wave length shifting fiber light transport (WLSF). Right) assembled detector.

Individual photo-electron signals from the photomultiplier tubes were amplified by a factor of ten in a fast amplifier and discriminated with a 0.5 photo-electron threshold. The resulting logic pulses with a width of 20 ns were digitized using a multi -channel scalar[22] with a clock period of 0.8 ns.  This allowed the summed number of photon pulses and coincidences between the two photomultipliers to be constructed in software. Because of the long mean decay time of the ZnS light emission, the summed number of resolved pulses provides an estimate of the energy deposited in the ZnS. A plot of the number of resolved pulses, labeled as photo-electrons (PE), is shown in Figure 4, for both UCN and background events. For the purpose of forming this plot, events during the holding time were considered background and events during counting were considered UCN events. A coincidence within 100 ns was required between the two phototubes to define the start of an event.  The total number of pulses from both tubes in an event was determined using a running integration gate.  Pulses in an event were integrated until no new pulse arrived for a time greater than a looking time parameter, equal to 4 us for the analysis presented in this paper.

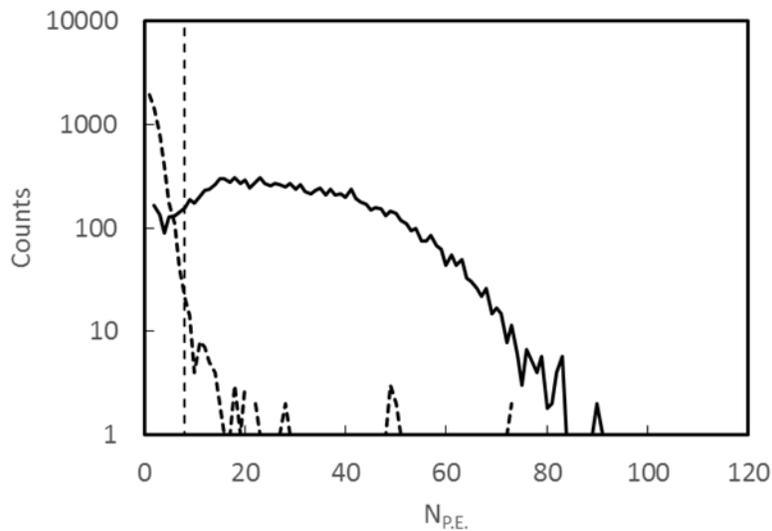

Figure 4) Spectra of the number of pulses (PE) detected for UCN+background events (solid line) and background events (dashed line). The vertical dashed line is PE=8, the threshold chosen to define a UCN event.

The pulse time distribution of the light from events in the detector was measured by creating a histogram of the pulses from a set of events as a function of time after the first pulse detected for each UCN+background event. The normalized results and a fit using three exponentials are shown in Figure 5. The fraction of the light in each exponential was, 0.18, 0.29, 0.53, and the time constants were, 0.134, 1.06, 5.90 μs, respectively. The bright short time constant provided sufficient statistics that a 100 ns coincidence width provided high-efficiency for counting UCN.

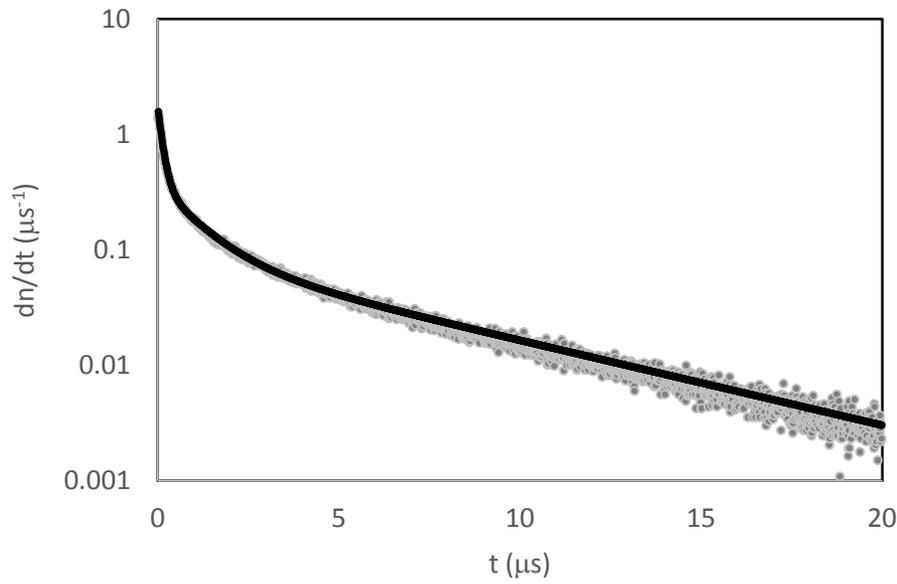

Figure 5) Detector pulse shape (grey points) and three exponential fit (black curve).

The efficiency of the dagger for light producing events was measured by mounting two 5-cm diameter phototubes above the trap with 7.5 cm diameter 7.5 cm focal length Fresnel lenses to image light from one side of the dagger onto the photomultiplier tubes.  A coincidence between these phototubes, DM, was used to tag UCN events from the adjacent side of the dagger.  The dagger efficiency was calculated using UCN events when the dagger was lowered in to the trap, the unloading peak, as:

$$eff = \frac{\text{Dagger} \bullet \text{DM}}{\text{DM}}, \quad\quad 3)$$

where the ● designates a coincidence and Dagger are UCN events detected by the dagger. The efficiency as a function of the threshold in terms of PE is shown in Figure 6. The efficiency for PE= 2, the minimum, is 0.976(2) and for PE= 8 the efficiency is 0.961(3).

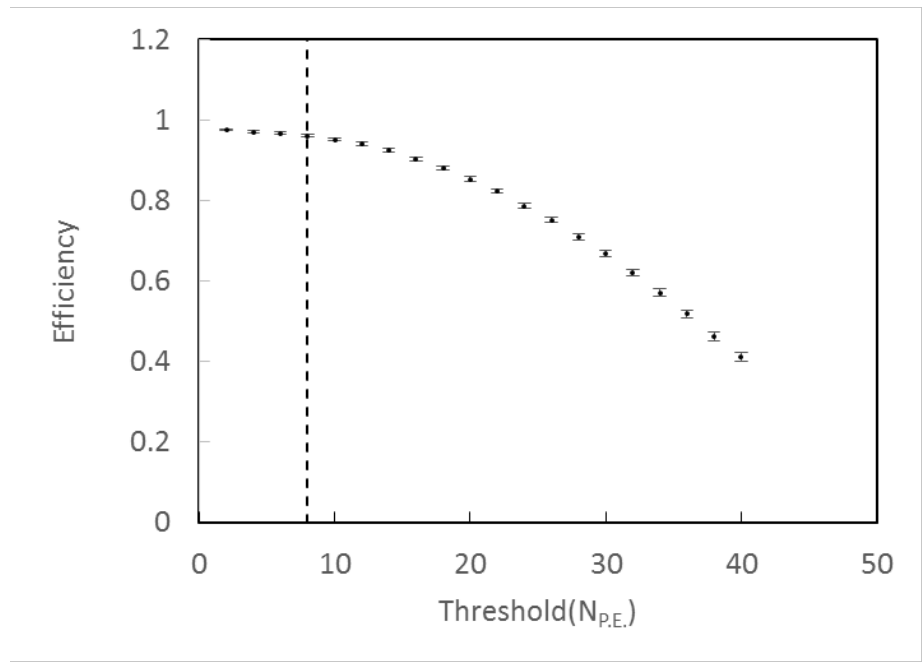

Figure 6) Efficiency as a function of the minimum number of PE required for an event. The dashed line is drawn at PE=8. From Figure 4 it can be seen that the higher threshold results in a large decrease in the sensitivity to background with little loss in efficiency.

## Cleaning

During commissioning, we made the first lifetime measurement with the dagger described above. The sequence for these measurements was to fill the trap for 150 s and clean the trap for 200 s, with the cleaner down for both operations. At the end of the cleaning time, the cleaner was raised for the storage and counting parts of the run cycle. At the end of the storage time, the dagger was lowered to 1 cm from the bottom of the trap and the remaining UCN were counted for 100 s. At the end of the counting time, the trap door was opened and post-counting remaining neutrons were drained into an *ex situ* detector, the Dump detector shown in Figure 2. No evidence of post-counting remaining neutrons was observed in the Dump detector.

The lifetime of neutrons in the trap was determined by counting the neutrons remaining after two different storage times. The initial number of neutrons loaded into the trap was determined by calculating yields normalized using two different monitor detectors (of five installed). The primary monitor, the standpipe detector (SP), was mounted at an elevation of 50 cm above the bottom of the trap on a tee in the UCN guide before the trap (see Figure 2), so it measured neutrons with energies above the maximum storable neutron energy of the trap. The second monitor detector, MON, was mounted near beam elevation and measured the incident UCN flux through an 8 mm diameter hole in the UCN guide near the biological shield wall. All of the monitors consisted of 10B-ZnS-PMT detectors described in Ref [21].

The loading time constant was approximately 60 s and the trap was loaded for 150 s to reach approximate saturation. The normalization was obtained by convolving the SP rate with an exponential with a time constant of 60 s with respect to the time the trap door was closed. The SP detector was chosen as the primary monitor because of its higher counting rate and better statistics. In addition, the energy spectrum of UCN from the source was found to harden slightly with beam exposure time, most likely due to beam radiation damage to the solid deuterium crystal. A linear correction which was a function of $T = MON/SP$ was applied to correct for these spectral changes. The yields were calculated as:

$$Y_{S,L} = \frac{N_{S,L}}{\int_{-150}^{0} SP(t) e^{\frac{t}{60}} dt} \left(1 + a \frac{T - \bar{T}}{\bar{T}}\right), \qquad 4)$$

where N is the raw number of detected neutrons remaining in the trap at the end of the storage time (the subscript S,L denote short and long holding times respectively), $a$ is a constant that was fitted to minimize the sum of root mean square (RMS) of the long and short yields for each set of runs, t=0 is the time relative to the time at which the trap door was closed and $\bar{T}$ is the average value of $T$ for the data set consisting of multiple S,L pairs of runs.

The lifetime of neutrons in the trap is then given by:

$$\tau = \frac{t_L - t_S}{\ln\left(\frac{Y_S}{Y_L}\right)}, \qquad 5)$$

where $t_L$, and $t_S$ are the long and short holding times respectively.

The first data set consisted of 45 short (20 s) and 32 long (1520 s) long holding time runs. The summed time distributions of UCN+background events from this data set are shown in Figure 7. The lifetime from these data was found to be 857.2±3.4 (stat.) seconds.

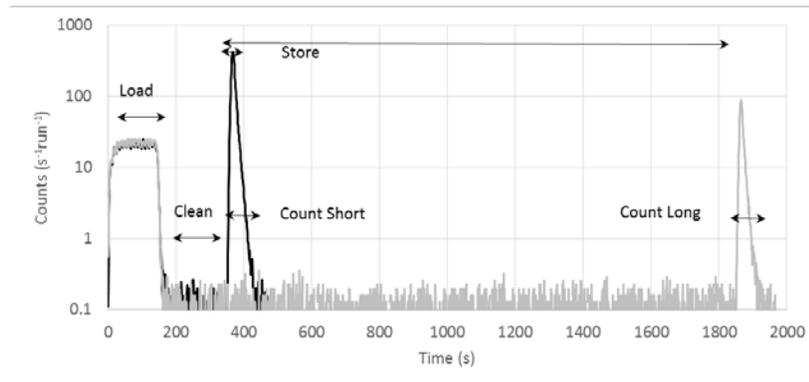

Figure 7) Summed short and long storage time time distributions from the first $\tau_n$ data set.

This resulted in the hypothesis that the UCN population in the trap was not sufficiently cleaned, and that quasi-bound neutrons were escaping during storage. In order to check this, the counting sequence was changed to lower the dagger in two steps, first to a position 37 cm from the bottom of the trap to count for 40 s (labeled 1 in Figure 7) and then to 1 cm from the bottom of the trap to count for for 60 s (labeled 2 in Figure 7) . The loading time and cleaning time for these data were 150 s and 300 s respectively. The normalized time distributions, shown in Figure 8, show a shorter lifetime for the first counting group than for the second. The lifetime extracted from the first, higher energy, group was 614±23 s, compared to 880±5 s after fitting and subtracting the remaining contribution from the first group. This contribution was calculated by fitting the first group data with an exponential in and correcting its time constant for the neutron lifetime to extrapolate the contribution of these neutron to the second counting group, assuming that these neutrons were counted with a short time constant in the lower dagger position. This measurement both showed the cleaning to be insufficient and provides a correction method.

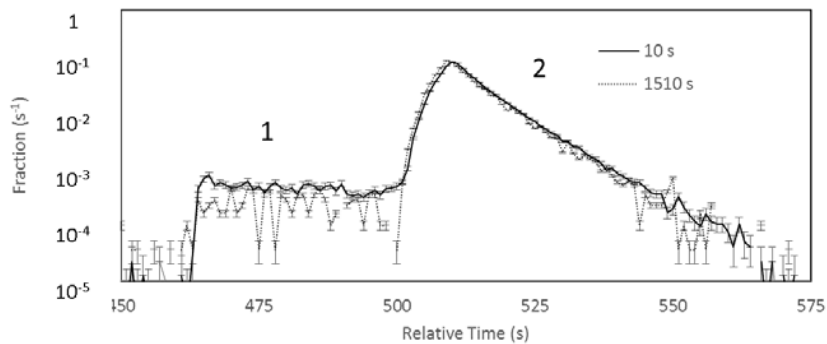

Figure 8) Overlay of the short and long, background subtracted, two step unloading time distributions for two step counting. The long holding time spectrum has been offset in time to line up with the short holding time spectrum, and both have been normalized by their integrals. Counting groups are labeled 1 and 2

A third measurement was performed by lowering the dagger to the upper counting position during the loading and cleaning (dagger cleaning), raising it entirely out of the trap during the storage period, and then doing two step counting of the remaining neutrons. The time distributions of long and short storage runs (normalized by the integral counts), overlaid in Figure 9, show fewer counts in the first counting group (1) by a factor of 2.3, and the ratio of long to short rates of the two groups are much closer to being equal. The lifetime from the first group is 779±49 and the corrected lifetime from the second group is 879.1±4.1, demonstrating more complete cleaning of quasi-bound neutron. All of these results are summarized in Table 2.

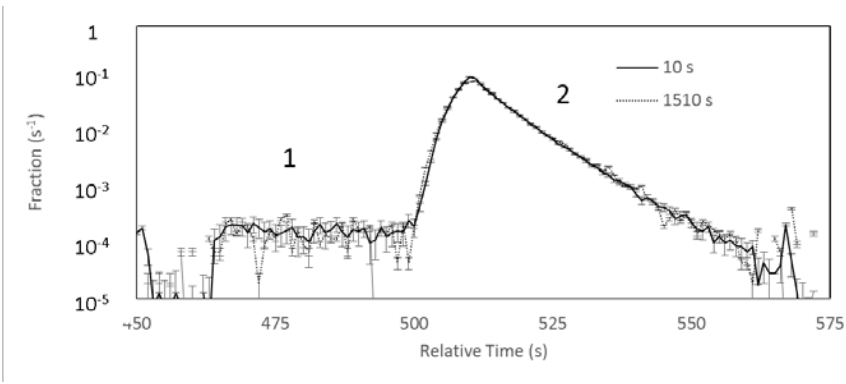

Figure 9) Same as Figure 8 but with dagger cleaning.

Although dagger cleaning reduced the lifetime correction for quasi-trapped neutrons to 1.8±0.8 s, this is still large compared to the goal of a 1 s counting statistics limited measurement. In order to further study energy distribution the trapped UCN a 4 step counting sequence was used, with dagger positions of 37, 25, 13, 1 cm from the bottom of the trap.

We have used four step counting to study other cleaning conditions. Since these data were part of production data taking, they were blinded, so lifetime results are not presented here. Some results are shown in Figure 10 to illustrate features of cleaning. The spectrum on the left shows a significant population of UCN in the counting group 1. The data on the right were cleaned using the same cleaner cycle as those on the left, but with the dagger lowered to 25 cm (well below the heigth of the lowered dagger of 37 cm) for the loading and cleaning times. In this mode there were negligible counts measured in the counting group 1, at a height of 37 cm. The relative number of counts in the counting group 2 is observed to increase with holding time. This is because the cleaning apparently reduced the population of neutrons in the region of phase space of orbits that are counted in the second dagger position. This creates a hole in the phase space which heals as neutrons redistribute in phase space at longer holding times. Because of the short counting times, this comparison demonstrates that the active dagger detector allows more effective probing of the dynamics of the trapped UCN than can be obtained in storage experiments experiments with traditional *ex situ* detectors.

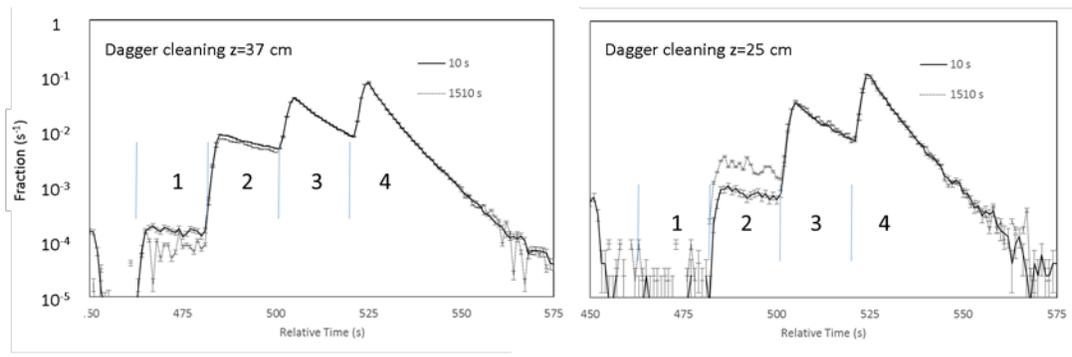

Figure 10) Overlay of the short (solid line) and long (dashed line), background subtracted four step unloading time distributions for 300 second cleaning using both the cleaner and the dagger, with the dagger lowered to 37 cm (left) and 25 cm (right). The numbers, 1-4, label the counting positions.

## Preliminary Lifetime Results

The neutron lifetimes obtained from the three two-step counting conditions described in the previous section are summarized in Table 2 and plotted in Figure 11. The live time corrected yields were calculated using equation 4. The data were analyzed in time adjacent pairs, and the lifetimes were calculated according to equation 5. Three corrections are applied to these lifetimes: first for the measured effect of uncleaned neutrons and second for the residual pressure in the trap. This correction was smaller for set A because the trapped neutrons are counted relatively 40 seconds sooner when compared to set B and fewer decay. The pressure correction was made using a calibrated cold cathode gauge to measure the pressure, an RGA to measure the mass spectrum of the gas and measured cross sections[23] to calculate the velocity independent UCN lifetime due to losses on the residual gas in the trap. Finally, the effects of control timing errors and phase space evolution were accounted for by using the centroid of the long and shot counting times to determine $t_l - t_s$. The corrections to the holding time were 2.5±0.5 s for data set A and B, dominated by control error, and 0.8±0.2 s for data set C dominated by phase space evolution. Here the uncertainties are statistical. Because of the short counting time, the phase space evolution correction and its uncertainty are small. Remaining systematic uncertainties are listed in Table 3.

One significant systematic uncertainty is due to dead-time/pile-up. The dead time correction to the short holding time runs is larger than for the long holding time runs. The dead time is calculated as the width the photon counting time is opened for each event. Monte Carlo simulations show that this slightly overestimates the dead time because an event within this gate can occasionally generate a coincidence after the end of the gate and be counted. We have estimated the size of the effect to be as large as 0.5s. The dead time algorithm will be improved in the future.

Table 2) Summary of measured and corrected lifetimes

| Set | Raw | | Cleaning | | Vacuum | | Corrected | |
|---|---|---|---|---|---|---|---|---|
| | $\tau_{measued}$ | $\Delta\tau_{measued}$ | $\tau_{correction}$ | $\Delta\tau_{correction}$ | $\tau_{correction}$ | $\Delta\tau_{correction}$ | $\tau_n$ | $\Delta\tau_n$ |
| | s | s | s | s | s | s | s | s |
| A | 858.4 | 3.5 | 18.2 | 1.8 | 0.4 | 0.1 | 877.0 | 4.0 |
| B | 862.8 | 5.7 | 17.4 | 1.5 | 1.6 | 0.5 | 881.8 | 6.0 |
| C | 876.5 | 4.0 | 1.9 | 0.9 | 0.9 | 0.3 | 879.3 | 4.1 |
| | | | | | | Average | 878.8 | 2.6 |
| | | | | | | $X^2$/dof | 0.24 | |

A. One step counting
B. Two step counting
C. Two step counting with dagger cleaning

The cleaning correction for the set of data taken using the dagger to augment the cleaner is relatively small (1.8 s) when compared to the sets where only the cleaner was used for cleaning. Never the less, the correction is observed to bring all three sets into statistical agreement, see Figure 11, supporting the conclusion that the dagger effectively measures the correction. Subsequent to these measurements a larger area cleaner was installed to improve the cleaning. In Table 3 we present estimates of the remaining systematic uncertainties.

Table 3) Estimated systematic uncertainties not included in Table 2)

| effect | upper bound (s) | direction | Current Eval. | Method of Characterization |
|---|---|---|---|---|
| depolarization | 0.01 | + | calculated | theory |
| microphonic heating | 0.1 | + | simulated | accelerometer studies |
| dead time/pileup | 0.5 | ± | simulated | coincidence studies |
| time dependent background | 0.1 | ± | measured | measurements |
| gain drifts | 0.2 | ± | measured | measurements |
| Phase space evolution | 0.2 | ± | measured | measurements |
| total | 0.6 | | (uncorrelated sum) | |

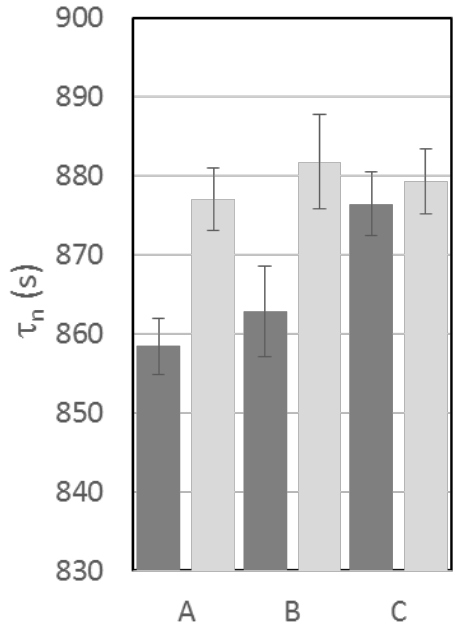

Figure 11) a plot showing the measured lifetimes (dark grey) and corrected for marginally trapped UCN (light grey) for three different sets of runs: A) one step counting, B) two step counting, C) two step counting with dagger cleaning.

## Conclusion

We have described a new method for *in-situ* counting of neutrons in a magneto-gravitational trap. The dagger detector allows the systematic correction for insufficient cleaning to be measured and the lifetime data to be corrected. The counting time using this detector is comparable to the uncertainty in the lifetime, ensuring that corrections due to phase space evolution on the neutron holding time can be measured to relatively high precision. Further, these measurements have led to the implementation of a more effective cleaner with a much larger surface area. This cleaner will be used in subsequent experimental campaigns.

Neutron lifetimes were extracted from three data sets that were taken using different cleaning conditions. These data sets resulted from our commissioning runs and were never blinded. The lifetimes extracted from the three sets of data are in agreement, and they give an average neutron lifetime $\tau_n$=878.8± 2.6± 0.6 s, in good agreement with previous bottle lifetime measurements but in disagreement with the beam measurements. The method described here will be applied to a blinded dataset with higher statistical sensitivity, and any remaining potential sources of systematic uncertainty will be quantified. Our definitive lifetime value will come from a future blinded analysis of later data sets.

LA-UR-16-27352

# Acknowledgements

This work was supported by the Los Alamos LDRD office, the Department of Energy, and the National Science Foundation (1307426). The authors would like to thank the staff of LANSCE for their diligent efforts to develop the diagnostics and new techniques required to provide beam for this experiment.